%% file: EAIndexArxiv.tex
\documentclass[12pt,a4paper]{scrartcl}
\usepackage[hidelinks]{hyperref}
\usepackage[latin1]{inputenc}
\usepackage{float,lscape}
\usepackage[ngerman, english]{babel}
\usepackage{lmodern}
\usepackage{graphicx}
\usepackage{geometry}
\usepackage{color}
\usepackage{xcolor}
\usepackage{anysize}
\usepackage{amssymb}
\usepackage{latexsym}
\usepackage{colortbl}
\usepackage{booktabs}
\usepackage{subfig}
\usepackage{multirow}
\usepackage{tikz}
\usepackage{amsmath}
\usepackage{jneurosci}
\usepackage{cite}

\usetikzlibrary{shapes,arrows,mindmap,trees,positioning}
\begin{document}
\newgeometry{left=20mm,right=20mm,top=2.5cm,bottom=2cm}
\title{Index of Environmental Awareness in Russia}\subtitle{{\LARGE MIMIC Approaches for Different Economic Situations}\footnote{We are very grateful to Yandex, which provided us with the necessary empirical data from Russian regions. Research was supported by RSF grant Nr. 15-18-20029 ``Projection of optimal socio-economic systems in turbulence of external and internal environment''. D. Khakimova wishes to acknowledge the support of the Ministry of Education and Science of the Russian Federation, grant No. 14.U04.31.0002, administered through the NES CSDSI.}}
\author{Dilya Khakimova\thanks{New Economic School, 100A Novaya Street, Skolkovo, Moscow, 143026, Russia. E-mail: d.a.khakimova@gmail.com} \and Stefanie Lösch\thanks{Dresden University of Technology, Faculty of Transportation, Chair of Statistics and Econometrics esp. Transportation, 01062 Dresden, Germany. E-mail: stefanie.loesch@tu-dresden.de} \and Danny Wende\thanks{Wissenschaftliches Institut für Gesundheitsökonomie und Gesundheitssystemforschung, WIG2 GmbH, Nikolaistr.\ 6-10, 04109 Leipzig, Germany. E-mail: wende@wig2.de} \and Hans Wiesmeth\thanks{Prof. em., Dresden University of Technology, Faculty of Business and Economics, Chair of Allocation Theory, 01062 Dresden, Germany; Ural Federal University, Graduate School of Economics and Management, 19 Mira Street, Ekaterinburg, 620002, Russia. E-mail: hans.wiesmeth@tu-dresden.de} \and Ostap Okhrin\thanks{Dresden University of Technology, Faculty of Transportation, Chair of Statistics and Econometrics esp. Transportation, 01062 Dresden, Germany. E-mail: ostap.okhrin@tu-dresden.de}}
\date{March 2017}

\def\R{I\!\!R}

\newcommand{\E}{\operatornamewithlimits{\mathbb{E}}}

%\pdfoutput=1
\baselineskip=16pt
\maketitle
\thispagestyle{empty}
\begin{abstract}
\noindent
\small{
The paper addresses the issue of environmental awareness in the regions of the Russian Federation. Russia provides an important study area to investigate the relationship between economic development and environmental consciousness in general. This paper introduces an index of environmental awareness, which is derived as a latent variable from various categories of search entries in Yandex, the prominent Russian search engine, during two periods in years 2014 and 2015. These indicators are presumably dependent on certain causes, which are also integrated into the model. The resulting Multiple Indicators-Multiple Causes model allows to estimate the proposed index of environmental awareness and to rank the Russian regions for each period. Comparing the results of 2014 and 2015 is especially interesting, because of the RUR devaluation at the end of 2014. 
In additional, we answer the question of the existence of an Environmental Kuznets Curve with respect to environmental understanding.} \\


\end{abstract}
%\vfill
%\pagebreak
\newpage
\restoregeometry
\section{Introduction}
\label{sec:introduction}
Environmental degradation remains a global issue and the Russian Federation is not an exception. Various cities and regions continue to suffer from poor environmental conditions due to or in spite of the economic decline after the breakdown of the Soviet Union and the economic recovery in recent years. In particular, air pollution is still responsible for unnecessarily high rates of morbidity and even mortality, see \cite{Golub2013}.

\begin{figure}[!h]
\centering
\includegraphics[trim=0mm 5mm 0mm 10mm, clip, width=0.6\textwidth]{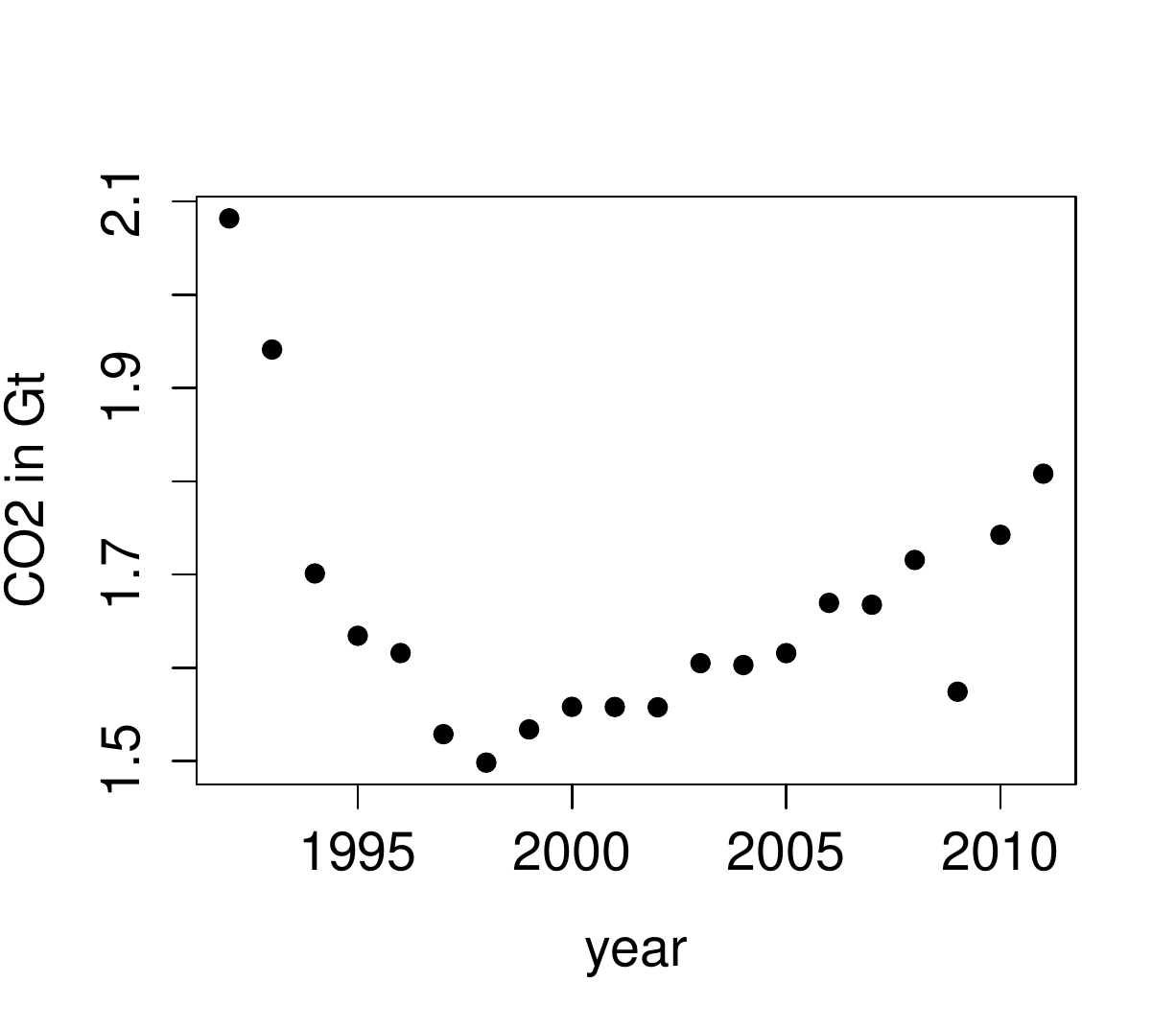}
\caption{CO$_2$ emissions (in Gt) of Russia from 1992-2011.}
\label{pic:co2}
\end{figure}

Fortunately, there are various signals, which document the political willingness to reduce environmental pollution. For example, in 1991 a system of fees on emissions, discharges, and solid waste became a key element of Russia's environmental policy. There were even differentiated fees per unit of emissions of hazardous substances, see \cite{Golub2013}, p. 436f. Moreover, in 2004 Russia was the last party to join the Kyoto Protocol, and was instrumental to its successful implementation. Thus, there was the readiness to participate in the provision of global environmental commodities, such as the reduction of greenhouse gas emissions. With emissions a little bit above 1.7 Gt CO$_2$ equivalents per year, as plotted on Figure \ref{pic:co2}, Russia is still among the largest five emitters of greenhouse gases after China, the United States, European Union and India. However, due to the economic recession following the breakdown of the Soviet Union, per capita emissions of greenhouse gases in Russia are still 24\% below their 1990 level, despite recent increases of 7\% in 2010 and 5\% in 2011, see \cite{IEA2013}.

It was relatively easy for Russia, to reach the Kyoto target of a 0\% reduction of GHG emissions in the period from 1990 to 2012. With joining the Kyoto Protocol, Russia could make use of this advantageous situation by offering their surplus emission reductions units, sometimes referred to as ``hot air'', to those states and companies in need of additional units. Nevertheless, in 2009 the Russian government passed the resolution ``On the Measures Stimulating Reduction of Atmospheric Pollution by Products of Associated Gas Flaring'', which limits associated gas flaring levels to 5\% of the entire output as from 2012, see \cite{IEA2013}. Resolutions like this certainly initiate substantial investments in clean technologies. 

Thus, in view of the current negotiations on global reductions of greenhouse gas emissions the question arises, to what extent people in Russian regions are environmentally aware. Russia is still one of the major emitter of greenhouse gases, and a growing environmental awareness (EA) can accelerate Russian efforts to mitigate climate change. 

\begin{figure}[!h]
\centering
\includegraphics[trim=0mm 5mm 0mm 10mm, clip, width=0.6\textwidth]{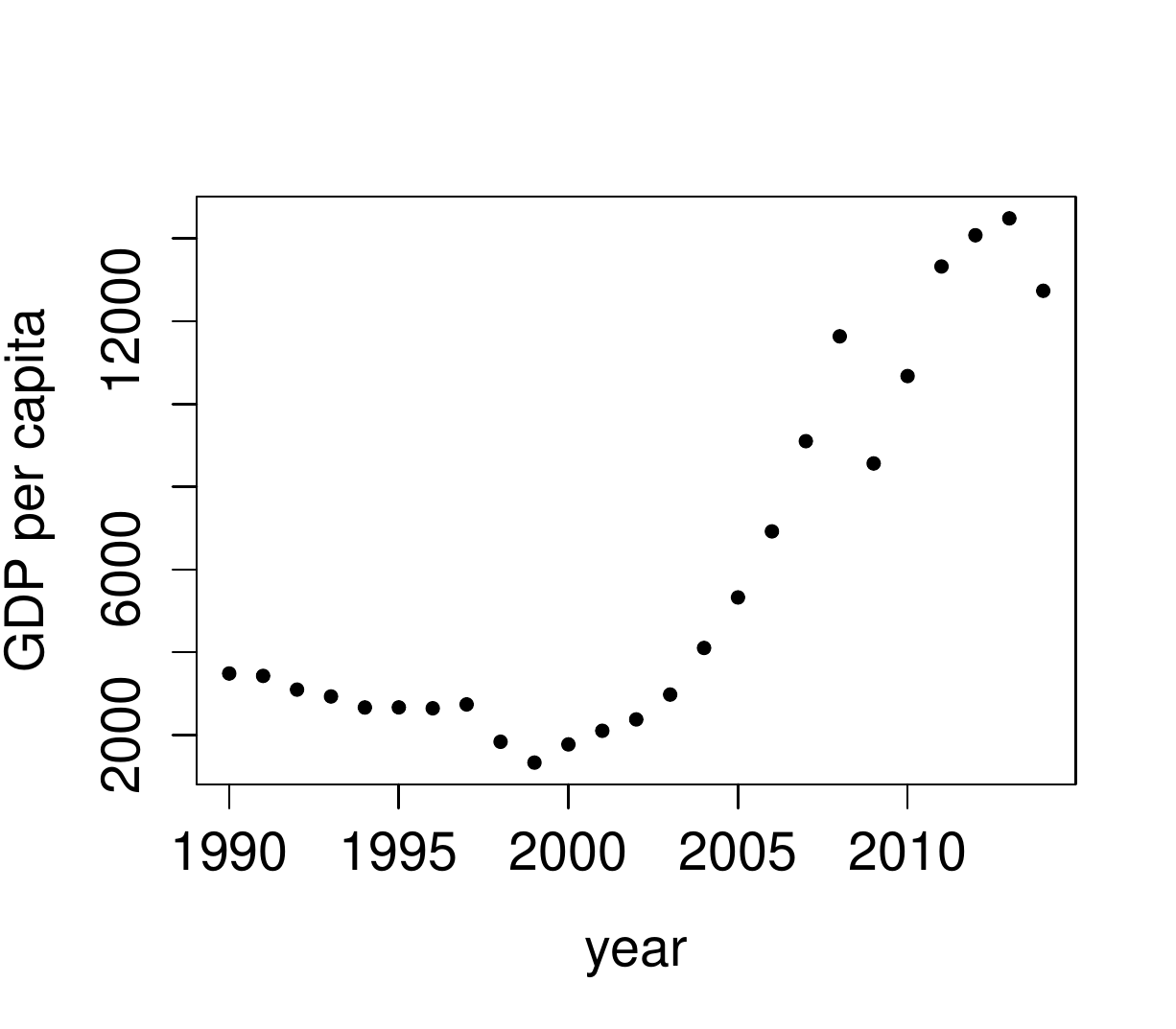}
\caption{GDP per capita in current USD of Russia from 1990-2014.}
\label{pic:gdp}
\end{figure}

The Russian Federation provides a particularly interesting study area for this kind of research. If there exists a relationship between economic development and whatever expression of EA, then environmentally friendly behaviour should play a more important role in Russia in general and the Russian regions in particular.

As Figure \ref{pic:gdp} shows, GDP per capita in Russia increased from 1\,771 USD in 2000 to 14\,611 USD in 2013, data source \cite{WorldB}. As there are some differences between the regions in terms of economic and demographic development and in terms of ethnic composition, we should expect differences regarding the emergence of EA in the regions of the Russian Federation.

The basic question therefore refers to the concept of EA. How can this awareness be measured under the conditions mentioned above? Moreover, what are the relevant differences among the Russian regions in this context? Which structural characteristics of the various regions are decisive for differences regarding EA? Is there a relationship between income and EA of a region? Does it have a polynomial relationship to a GDP per capita? In other words does it look similar to the so-called Environmental Kuznets Curve (EKC)? 

Many empirical studies investigate the direct effects of economic, demographic and political conditions on the environment and derive there from political measurements. However, most of the time they do not consider the sentiments or understanding of environmental relevant issues of the population. Certainly, the latter is important to perform environmental policies successfully.
For this reason, we develop a new index of EA for Russia and provide a ranking for 81 Russian regions for two periods in years 2014 and 2015.
This index of EA is a latent construct, which affects various indicators. These indicators are derived from search entries in Yandex, the prominent Russian search engine. The data refers to some 200 phrases, both in Russian and in English, related to environmental issues. These search phrases are clustered into five categories covering, among others, direct climate change queries as well as literature searches.

We assume that these indicators depend on or, rather, result from certain economic activities or societal developments in the various Russian regions. Thus, we integrate these causes into the model to explain a specific level of the index. The resulting model is the Multiple Indicators-Multiple Causes (MIMIC) model, which allows to estimate the latent variable from various indicators and various causes of EA, see \cite{Buehn2013}.

We also look for differences in the EA between 2014 and 2015. We except a smaller interest in environmental topics in 2015 than in 2014, because of the devaluation of the Russian currency RUR at the end of 2014.

The paper is structured as follows. Section \ref{sec:introduction} introduces. In section \ref{sec:Literature} we briefly investigate the development of the concept of EA in the literature starting with the marketing and social psychology literature some 50 years ago. The formal aspects of the MIMIC model are presented in section \ref{sec:MIMIC} including a discussion on the applicability of this model to the concept of EA. The next step is the introduction of indicators and potential causes of environmental pollution or EA. First results of our investigations are given in section \ref{sec:Results}. Section \ref{sec:Concluding} concludes.

\section{Environmental Awareness in the Literature}
\label{sec:Literature}
\subsection{Marketing and Social Psychology}
Almost everybody understands EA as some kind of environmentally friendly behaviour, however it is not so straightforward to conceptualize it in order to make stringent use of it in academic research.

Not too surprisingly, interest in the concept of EA or environmental consciousness originated with the ecological movement in the 1960s. According to \cite{Soyez2009}, researchers in marketing and social psychology focused first on personal characteristics of environmentally conscious people, such as sociodemographic variables. In the 1970s and 1980s environmentally friendly behaviour was more explained in terms of environmentally friendly attitudes measurable by means of multi-item scales. The ``theory of planned behaviour'', applicable to this context (see \cite{Ajzen1991}) allows an integration of a variety of factors or attitudes (self interest, norms, situational barriers, etc.) that influence a specific action or behaviour related, for example, to the environment. 

Personal value orientation as precursor of sustainable behaviour was considered in a further stream of research followed by cultural values, which have been investigated approximately for the last ten years, see \cite{Soyez2009}. Of course, cultural values form the basis for cross-cultural studies on environmentally friendly behaviour, which are of particular interest for researchers in marketing and social psychology. In this context \cite{Soyez2012} analyses how environmentally friendly behaviour is influenced by cultural values, how national cultural values can be linked to personal pro-environmental behaviour. This study is among the first to provide insight into environmentally friendly consumer behaviour in Russia.

\subsection{The Environmental Kuznets Curve}
It is natural to assume that environmental commodities are characterized by a relatively high income elasticity, at least in industrialized and newly industrialized countries. Consequently, demand for these commodities should rise, and environmental pollution should be reduced with real GDP per capita increasing. The location of the maximum of the resulting functional relationship between GDP per capita and the level of pollution is described by the EKC. It could be used as an indicator of a gradually emerging EA: a further increasing GDP per capita will be accompanied by increasing efforts to reduce the pollution. 

In this context, \cite{Grossman1995} find no evidence that environmental quality deteriorates steadily with economic growth. In their words, for most indicators, economic growth brings an initial phase of deterioration followed by a subsequent phase of improvement. The turning points for the different pollutants vary, but in most cases they come before a country reaches a per capita income of \$8000 (for 1985). \cite{Grossman1995} use urban air pollution, the state of the oxygen regime in river basins, fecal contamination of river basins, and contamination of river basins by heavy metals. 

The EKC is thus a hypothesized relationship between various indicators of environmental pollution and GDP per capita, see \cite{Stern2004}. The concept emerged in the early 1990s with studies of the potential environmental impacts of NAFTA. \cite{Stern2004} provides an interesting survey on the rise and the fall of the EKC, characterizing the EKC as an essentially empirical phenomenon, with not much support from econometrics.
Similarly, \cite{Huang2008} show that there seems to be no empirical evidence supporting the EKC hypothesis for greenhouse gas emissions. 

Nevertheless, in the last years more and more advanced econometric techniques were employed to investigate the existence or non-existence of the EKC with respect to aspects of global warming. \cite{Fosten2012}, for example, analyze the EKC with respect to CO$_2$ and SO$_2$ emissions in the UK, and provide a useful literature survey on the econometric methods used in this context (see \cite{Brajer2011}; \cite{He2010}; \cite{Wang2013}; \cite{Yang2015}).
\cite{maddison2006} shows that the countries' emissions per capita are affected by events in neighboring states. These effects should be considered creating a EKC. 
 A comprehensive survey of the EKC hypothesis is provided by \cite{Dinda2004}.

Already \cite{cassou2004} investigate the relationship between economic growth and environmental degradation, which are significantly influenced through the fiscal policy of a region.  

Recently, other empirical investigations revealed interesting aspects of the willingness to pay for climate actions. In this context \cite{Diederich2014} uncover causes of voluntary climate action, among them education, the information structure among the population, and exogenous environmental conditions. In a similar way, \cite{Borick2011}, and \cite{Lorenzoni2006} study public views on climate change in the US and Canada, and in Europe and the US, respectively. The focus of the special survey ``European's attitude towards climate change'' in the ``Eurobarometer'' from the \cite{EC2008} was on, among other issues, the extent to which citizens feel informed about climate change, thus on ``awareness of climate change''. Awareness regarding climate change has also been addressed in various publications. \cite{Zyadin2014}, for example, investigate the perceptions regarding renewable energies of senior academics and early-stage researchers involved in renewable energy sciences. Similarly, \cite{Karytsas2014} examine the demographic and socioeconomic factors that determine someone's knowledge on different forms of renewable energy. Moreover, \cite{Weber2015} address the issue of EA by means of a theoretical model involving the Nash mechanism.

\subsection{Indices of Environmental Quality}
Besides behavioural characteristics and the EKC hypothesis, indices of environmental quality have been introduced to categorize EA. There is, for example, the Environmental Performance Index (EPI), which measures the environmental performance of a country or a region according to a list of parameters. This list includes among others: air pollution, water, biodiversity and fisheries, see \cite{Buehn2013}.

The problem with this approach to EA is that changes in data sources, the weighting of the categories and the aggregation method make it difficult to compare the index for different years. In a situation with EA gradually emerging, it might not be a good idea to use performance indicators with fixed weights for the indicators. For this reason, we rather prefer the MIMIC approach with endogenously assigned weights. 

\subsection{Multiple Indicators and Multiple Causes}
Given these not too satisfactory results for appropriate indicators of EA, we pick up the approach used by \cite{Buehn2013} for an index of air pollution. Their index is based on a MIMIC model making use of the fact that there typically is no single indicator of air pollution and that multiple causes affect air pollution. Similarly, EA can hardly be described by means of a single indicator, and different framework conditions might lead to different levels of awareness.

As already mentioned above, this approach helps to overcome the difficulty with a priori fixed weights for the various indicators in a situation where awareness is still emerging. This might currently characterize the situation in the Russian Federation and the Russian regions.

\section{The Multiple Indicators-Multiple Causes Model}
\label{sec:MIMIC}
\subsection{The Method}
\label{sec:method}
The MIMIC model was developed originally by \cite{Joereskog1975} and is a special case of the general structural equation model. It uses well defined indicators to measure a latent construct (index of EA) with associated properties and regresses them against theoretically discovered causes. Lately \cite{Buehn2013} use the MIMIC approach to construct a new index of air pollution for 122 countries for the period between 1985 and 2005 and discuss the advantages of this technique. 

The two parts of the model can be explained as a measurement model for the latent construct and a structural part, which describes the causal structure of the model. The measurement part can be described as follows:
\begin{align}
y=\lambda\,\eta + \varepsilon,
\label{eqn_y}
\end{align}   
where $y=(y_1,y_2,\dots, y_p)^\top$ is a set of observable endogenous indicators, which are affected by the EA, which is the latent variable $\eta$, and ${\varepsilon}=(\varepsilon_1,\varepsilon_2,\dots, \varepsilon_p)^\top$ being a vector of $p$ random errors. The factor loadings are summarized in the $p$-vector $\lambda$. The structural part follows our theoretical assumptions and can be written as
\begin{align}
\eta=\beta^\top{x}+\zeta,
\label{eqn_eta}
\end{align}  
where ${x}=(x_1,x_2\dots,x_k)^\top$ are exogenous causes, ${\beta}=(\beta_1,\beta_2,\dots, \beta_k)^\top$ is a set of model parameters, and $\zeta$ being a random error term.
Inserting (\ref{eqn_eta}) into (\ref{eqn_y}) results in
\begin{align}
{y} &=  {\lambda}\,\left({\beta}^\top{x}+\zeta\right) +{\varepsilon}={\Pi}^\top{x}+{v}, \label{inequation}
\end{align}
with ${\Pi}={\beta}\,{\lambda}^\top \text{ and } {v}={\lambda}\,{\zeta}+\varepsilon$.
It is assumed that the random errors ${\varepsilon}$ and $\zeta$ are mutually independent and normally distributed
\begin{align}
{\varepsilon}\sim \operatorname{N}({0},{\Theta}^2) \hspace{0.5cm} \text{ and } \hspace{0.5cm} \zeta\sim \operatorname{N}(0,\sigma^2),
\end{align}
with $\operatorname{E}(\zeta{\varepsilon}^\top)=0$ where 
${\Theta}=\operatorname{diag}(\theta_1,\theta_2,\dots,\theta_p)$. The parameters of $\lambda$ and $\beta$, and the variances $\theta^2$ and $\sigma^2$ of the error terms $\varepsilon$ and $\zeta$, can be estimated using a ML approach. 

There is indeterminacy in the structural parameters (if $\lambda$ is multiplied by a scalar and $\beta$ and $\sigma^2$ are divided by the same scalar parameters do not change). Avoiding this indeterminacy, we set one parameter fixed, see \cite{Goldberger1971}. 
Following \cite{Joereskog1975} the fixed case, where $x$ is fixed and $y$ is normally distributed, $y \sim \operatorname{N}(\Pi^\top{x},\Omega)$, the log-likelihood can be computed through
\begin{align}
\mathcal{L}(\beta,\,\lambda,\,\theta^2,\,\sigma^2)=-\frac{N}{2}\{\operatorname{log}|{\Omega}|+\operatorname{tr}({\Omega}^{-1}{W})\}-\frac{p\,N}{2}\log(2\pi)\rightarrow \text{ max!}
\end{align}
where $N$ is the sample size, ${\Omega}=\operatorname{E}\left({v}{v}^\top\right) =\sigma^2{\lambda}{\lambda}^\top+{\Theta}^2$ and ${W}=({y}-{x}{\Pi})^\top({y}-{x}{\Pi})$.
The parameters $\lambda$, $\beta$ and variances $\theta^2$, $\sigma^2$ can be estimated by setting the first partial derivative of $\mathcal{L}(\beta,\,\lambda,\,\theta^2,\,\sigma^2)$ with respect to all arguments to zero. 

The normality is a strong assumption, which leads to consistent, but inefficient estimators. The variances of the parameters cannot be exactly determined. This leads to more inexact indication of the parameters significance. For that reason, we use two different methods to correct the estimated variances and get two kinds of estimated parameter variances and thus two different standard errors.
The first method of correction comes from \cite{Satorra1994}. Thereby, the incorrect standard errors are weighted at the asympotic covariance-matrix (hereafter referred to as {\itshape MLM}).
Another approach uses a kind of sandwich estimator correcting the standard errors (hereafter referred to as {\itshape MLR}) from \cite{Yuan2000}. 
Determining the parameter significance, we look after the largest corrected standard error. 

\subsection{Data and Empirical Model}
In our model the indicators $y$ are queries of relevant environmental phrases from the internet search engine Yandex,  in English or in Russian. Getting different indicator variables in terms of different environmental topics, we cluster these phrases into following categories:
\begin{itemize}
\item[Y1:] Climate Change Queries, 
\item[Y2:] Endangered Environment Queries,
\item[Y3:] Political Queries,
\item[Y4:] Science Queries,
\item[Y5:] Renewable Energies and Technologies Queries,
\end{itemize}
While each phrase is characterised  by a specific field of interest concerning EA, most suitable notations were counted too. Following the data collection, we summarized compatible requests in each region and on each category and divide them by the sum of all requests from Yandex which had happened in the same regions. The indicator variables $y$ is thus computed as follows,
\begin{align}
y_{in} &=\frac{\text{number of queries of category $i$ in region $n$}}{\text{number of all queries in region $n$}},
\end{align}
where $i=1,\dots, p$ with $p=5$ is the index of category and $n=1,\dots, N$ the index of Russian region with $N=81$.
For each region $n$ we describe the measurement models in terms of the EA $\eta$ as follows:
\begin{align} 
\begin{pmatrix}
\textmd{y}_1\\ 
\textmd{y}_2 \\ 
\textmd{y}_3 \\ 
\textmd{y}_4 \\ 
\textmd{y}_5 \\ 
\end{pmatrix} =
\begin{pmatrix}
1 \\ 
\lambda_2 \\ 
\lambda_3  \\ 
\lambda_4  \\ 
\lambda_5  \\ 
\end{pmatrix} \times \eta + 
\begin{pmatrix}
\varepsilon_1 \\ 
\varepsilon_2 \\ 
\varepsilon_3  \\ 
\varepsilon_4  \\ 
\varepsilon_5  \\ 
\end{pmatrix}.
\label{mesuere}
\end{align}
In addition, explaining the index variable $\eta$, observable causes are needed. The gross regional product (GRP) per capita in purchasing power parity in first, second and third order is considered to allow for the EKC. Furthermore, we control for the structure of the industry (mining and manufacturing), the emissions of greenhouse gases per capita and some controls for agglomeration, social status and education.
From (\ref{inequation}) it follows:

\begin{align}
\begin{pmatrix}
{y_1}\\ {y_2}\\ {y_3}\\  {y_4}\\  {y_5}\\ 
\end{pmatrix}
=\begin{pmatrix}{\lambda_1}\\ {\lambda_2}\\ {\lambda_3}\\  {\lambda_4}\\  {\lambda_5}\\ 
\end{pmatrix}
\underbrace{\left[ \begin{pmatrix}
{\beta_1}\\ \vdots\\  \beta_k
\end{pmatrix}^\top
\begin{pmatrix}
\text{GRP per capita} \\ \text{GRP per capita}^2 \\ \text{GRP per capita}^3 \\ \text{mining} \\ \text{manufactoring} \\ \text{emissions pc} \\ \text{set of control variables} \\
\end{pmatrix} + \zeta\right]}_{\eta=\text{Index of Environmental Awareness}} +
 \begin{pmatrix}
\varepsilon_1\\ \varepsilon_2\\ \varepsilon_3\\ \varepsilon_4\\ \varepsilon_5\\ \end{pmatrix}.
\end{align}

\begin{figure}[!ht]
 	\centering
 	\includegraphics[trim=35mm 115mm 29mm 33mm, clip, width=0.7\textwidth]{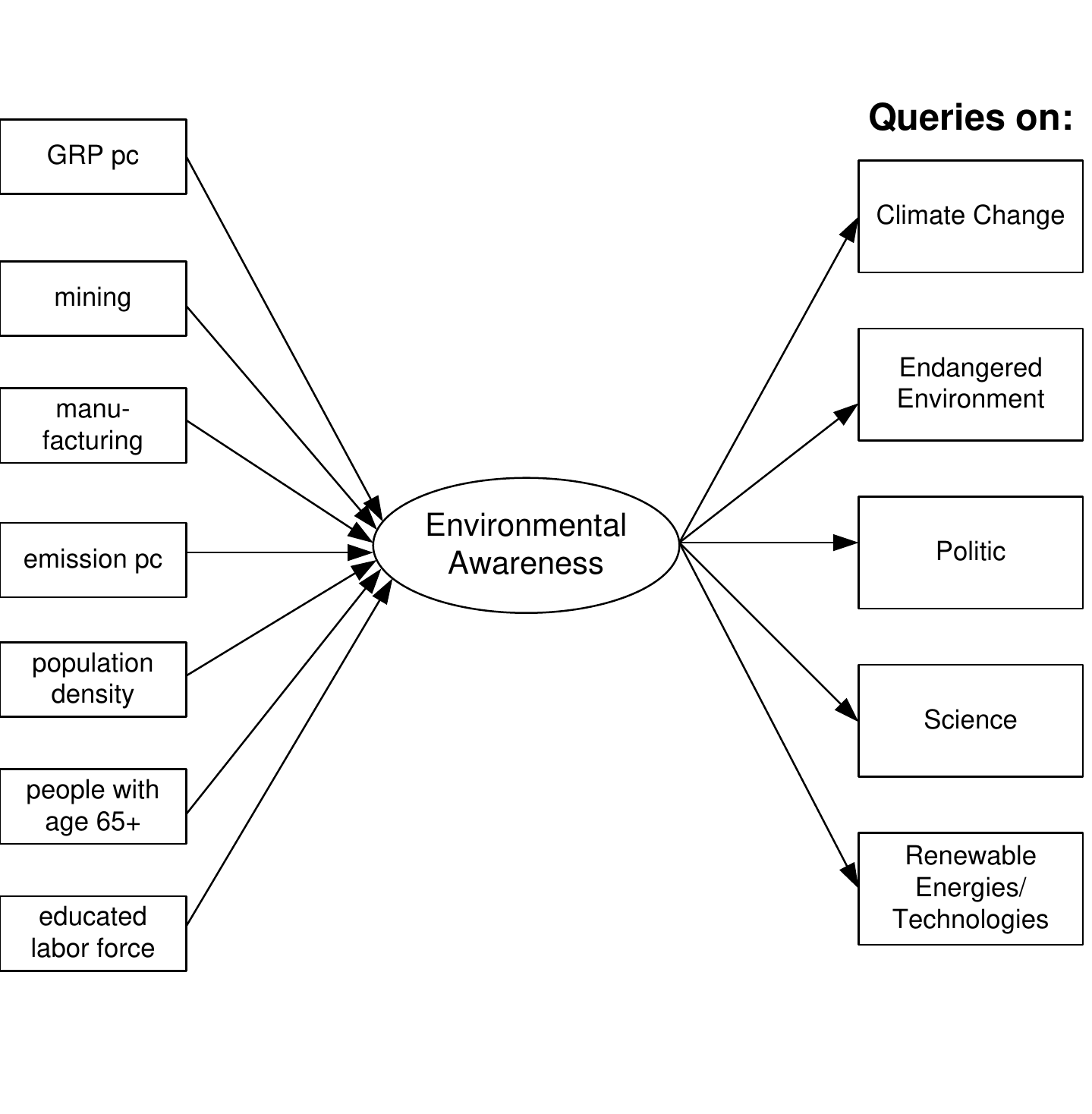}
 	\caption{Path Diagram of the MIMIC Model}
 	\label{pathdiagram}
 \end{figure}

Figure \ref{pathdiagram} shows the model with all specifications. Arrows mark the direct effects of the exogenous variables on the EA, and the mediation effect of the index on the indicators. 

With the obtained results, the score of the environmental EA $\eta$ can be calculated for each region. The EA thus provide a ranking of awareness of climate change in the regions of the Russian Federation. 
The data was collected from various sources, mainly \cite{OECD}, \cite{UNDP2013} and the \cite{RFSSS}. 
The gross regional product per capita in purchasing power parity, greenhouse gas emission per capita, population density, age structure and education of labour force come from the Organisation for Economic Co-operation and Development (OECD). 
The industry variables are from the Federal Statistics Service of Russia (RFSSS). The share of the mining and manufacturing industry was measured in terms of gross regional produced.

\subsection{Identification and Data Considerations}

As mentioned above, avoiding indeterminacy in the MIMIC-model, we set $\lambda_1=1$ fixed for the estimation. Making the values of the $\lambda$-parameter comparable and interpretable, all $\lambda_i$ are later standardized to have $\sum_{i=1}^p\lambda_i=1$. 
 
The data should follow the required distribution while using a ML-approach to estimate the parameter values of $\lambda$, $\beta$ and the variances of the error terms $\theta^2$ and $\sigma^2$. We performed Mardia's multivariate normality test, see \cite{Trujillo-Ortiz2003}, which however was rejected with high significance level. Consequently, we use two different methods to correct the estimated variances as described in section \ref{sec:method}.

\section{Results}
\label{sec:Results}
\subsection{The Results of the MIMIC model}
The MIMIC model was estimated with all specifications discussed in the previous section for July, September, November 2014 and January, March and May 2015. We aggregate the data for each year. 
It is important to compare the periods at the end of 2014 and at the beginning of 2015, because of the changes in the political situation in this time. Between 2014 and 2015 Russian currency RUR sharply devalued. Following the EKC we expect that the Russian population is less interested in environmental relevant topics in the second period than in the first.
To avoid bad scaling of the empirical covariance matrix and therefore resulting problems in the process of convergence, all included variables are standardised. 
The resulting estimates are given in Table~\ref{tab:results}.

We differ between two model patterns for each period: (A) and (B). Model (A) and (B) include the same query groups as indicators. Model (A) considers three more cause variables than model (B): $k=9$ cause variables for model (A) and $k=6$ for model (B). The reason is the variables mining, share of people older than 65 years and educated labour forces are  significant in some periods, but they deteriorate model fit parameters like Akaike information criterion (AIC), Bayesian information criterion (BIC), Comparative Fit Index (CFI), Root Mean Square Error of Approximation (RMSEA) and Standardized Root Mean Square Residual (SRMR). Small values for AIC, BIC and RMSEA and SRMR show a good model fit.

The indicators of EA have a positive sign, as expected, or otherwise are not significant, as the political and science queries in 2015. In general, the indicators are positively correlated with the new index, what supports our intuition that the index stands for high interests in the environment. 
Furthermore, there are some differences between 2014 and 2015 in the parameter values, perhaps as result of the RUR devaluation at the end of 2014. Unfortunately, these differences are not significant. Possibly, there is a latency of the effect, which can be measured if we include further queries from the second half year of 2015.

The first variable, the gross regional product per capita indicates a strong relation between regional wealth and the EA. This relation seems to be curvilinear, because of the significance of the GRP per capita parameter in the second and third order. 
The mining variable, which controls for the importance of the mining and oil industry in each region, shows a negative sign as the amount of this sector increases. But the standard errors are too high as being significant. 
More manufacturing also implies smaller EA.
A population with lots old people is less interested in terms of the future and therefore in information on the environment in contrast to the possibilities of the internet. More educated people, for instance a high share of labour force with tertiary education, lead to higher interests in environmental topics. If we instead include other attributes of education in our model (for example literacy rate) it leads to the same results. 

All in all, there are small differences between the parameter values of the cause variables between model (A) and model (B) and also from period to period. The signs however are kept the same.

\thispagestyle{empty}
\begin{table}[htbp]
\centering
\setlength{\tabcolsep}{0.1em}
\footnotesize
\caption{Estimated parameters for 2014 and 2015 using MIMIC approach}
\label{tab:results}
\begin{tabular}{lrp{0.7cm}rp{0.7cm}rp{0.7cm}rp{0.7cm}}\toprule
Parameter		&	2014	&		&	2014 	&		&	2015	&		&	2015	&		\\	
		&	(A)	&		&	(B)	&		&	(A)	&		&	(B)	&		\\	\midrule
$\lambda$		&		&		&		&		&		&		&		&		\\	
		&		&		&		&		&		&		&		&		\\	
Climate Change		&	0.196	&		&	0.198	&		&	0.356	&		&	0.383	&		\\	
		&		&		&		&		&		&		&		&		\\	
Endangered Environment		&	0.161	&	**	&	0.165	&	**	&	0.190	&	**	&	0.212	&	**	\\		
		&	(0.078)	&		&	(0.079)	&		&	(0.092)	&		&	(0.098)	&		\\	
Politic Queries		&	0.194	&	*	&	0.195	&	*	&	0.062	&		&	0.055	&		\\	
		&	(0.101)	&		&	(0.104)	&		&	(0.057)	&		&	(0.055)	&		\\	
Science Queries		&	0.237	&	***	&	0.237	&	***	&	0.045	&		&	0.045	&		\\	
		&	(0.078)	&		&	(0.081)	&		&	(0.121)	&		&	(0.114)	&		\\	
Renewable Energies/Technologies		&	0.212	&	**	&	0.206	&	**	&	0.347	&	***	&	0.305	&	***	\\	
		&	(0.102)	&		&	(0.105)	&		&	(0.091)	&		&	(0.075)	&		\\	
$\beta$		&		&		&		&		&		&		&		&		\\	
		&		&		&		&		&		&		&		&		\\	
GDP per Capita in ppp		&	2.618	&	***	&	2.411	&	***	&	3.483	&	***	&	3.358	&	***	\\		
		&	(0.969)	&		&	(0.878)	&		&	(1.111)	&		&	(1.067)	&		\\	
GDP per Capita$^2$ in ppp		&	-6.671	&	***	&	-5.546	&	***	&	-8.457	&	***	&	-7.335	&	***	\\		
		&	(2.329)	&		&	(2.046)	&		&	(2.592)	&		&	(2.504)	&		\\	
GDP per Capita$3$ in ppp		&	4.513	&	***	&	3.617	&	***	&	5.438	&	***	&	4.526	&	***	\\		
		&	(1.564)	&		&	(1.360)	&		&	(1.655)	&		&	(1.604)	&		\\	
Mining		&	-0.100	&		&		&		&	-0.169	&		&		&		\\		
		&	(0.153)	&		&		&		&	(0.200)	&		&		&		\\	
Manufacturing		&	-0.149	&	**	&	-0.287	&	***	&	-0.208	&	***	&	-0.325	&	***	\\	
		&	(0.072)	&		&	(0.081)	&		&	(0.068)	&		&	(0.106)	&		\\		
Emission per Capita		&	-0.291	&	***	&	-0.281	&	**	&	-0.322	&	***	&	-0.319	&	**	\\		
		&	(0.101)	&		&	(0.109)	&		&	(0.108)	&		&	(0.139)	&		\\	
Population Density		&	-0.247	&	**	&	-0.264	&	***	&	-0.334	&	***	&	-0.364	&	***	\\		
		&	(0.115)	&		&	(0.101)	&		&	(0.104)	&		&	(0.101)	&		\\	
People with Age 65+		&	-0.375	&	***	&		&		&	-0.437	&	***	&		&		\\		
		&	(0.113)	&		&		&		&	(0.151)	&		&		&		\\	
Labour Force with Tertiary Education		&	0.156	&		&		&		&	0.253	&	**	&		&		\\		
		&	(0.095)	&		&		&		&	(0.099)	&		&		&		\\	
		&		&		&		&		&		&		&		&		\\	\midrule
Number of Observation		&	81	&		&	81	&		&	81	&		&	81	&		\\	
AIC		&	2158.7	&		&	1699.8	&		&	2245.8	&		&	1788.4	&		\\	
BIC		&	2204.2	&		&	1738.1	&		&	2291.3	&		&	1826.7	&		\\	
CFI		&	0.726	&		&	0.760	&		&	0.814	&		&	0.852	&		\\	
RMSEA		&	0.081	&		&	0.079	&		&	0.064	&		&	0.060	&		\\	
SRMR		&	0.075	&		&	0.085	&		&	0.083	&		&	0.086	&		\\	
 \toprule
\multicolumn{9}{l}{\scriptsize Standard errors in parentheses; significance level: ***$p<0.01$, **$p<0.05$, *$p<0.1$}
\end{tabular}
\vspace*{-0.1cm}
%\end{table}
\end{table}

\newpage

\subsection{A Ranking of Environmental Awareness in Russia}
The construction of the index of EA gives us some new insight into the environmental situation of the Russian regions. As all variables in our model are standardised, the index is standardised as well and multiplied by 100 for better clarity, and computed separately for the years 2014 and 2015. 

\begin{figure}[!ht]
	\centering
	\includegraphics[trim = 0mm 0mm 0mm 20mm, clip, width=0.5\textwidth]{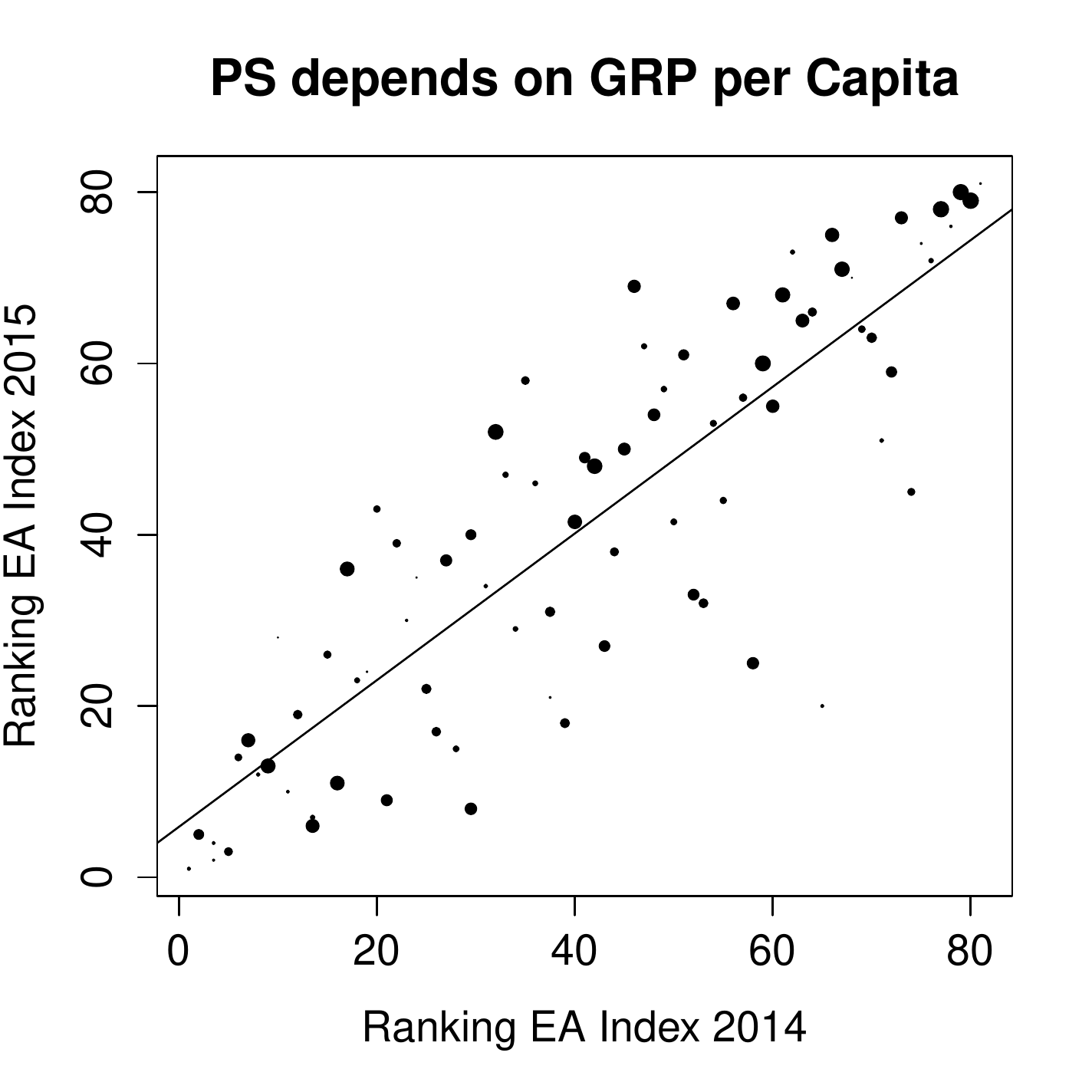}
	\caption{Relationship between the Ranking of EA Index in 2014 and in 2015. Point size depends on GRP per capita.}
	\label{fig:indexplot}
\end{figure}

As shown in Figure \ref{fig:indexplot}, there is a linear relationship between the EA Index in 2014 and 2015. The results hardly change from period to period and seem to be robust. If a region is highly ranked in 2014 it is in the most cases good placed in 2015, as well.

Interpreting the EA Index, we are looking for a connection between the gross regional product per capita and the EA. Indeed, if the income goes up our EA Index increases - mainly in a range between 14\,000 and 19\,000 USD.  
The largest regions with a corresponding GRP per capita show a high EA. Only the first placed region, Karbardino-Balkaria Republic, which is located in the North Caucasus, is an exception. The local economy is focussed on agriculture, forestry and mining, but also tourism plays an important role. The breakdown of the Soviet Union and further conflicts have led to an enormous decrease of tourism and the degradation of the regional economy. Although, the GRP per capita in 2013 amounts less than 8000 USD per capita, the EA Index reaches the highest value.
In contrast, the second and third placed regions, Nenets Autonomous Okrug and Khanty-Mansi Autonomous Okrug, which live mainly from mining and oil production, are among the richest regions of the Russian Federation. 

On the other hand, there are poorer regions like Krai Altai, Kemerovo and Kirov with the three worst indices in 2014 and Voronezh, Ivanovo and  Kirov in 2015.
These regions have an above average share of people older than 65 years and a high share of manufacturing industry.

The computed EA Indices for 2014 and 2015 are illustrated as maps in Figure \ref{fig:map2014} and Figure \ref{fig:map2015} in the Appendix.

\begin{table}[htbp]
\footnotesize{\sffamily
      \caption{The Environmental Awareness Index for 81 Russian Regions for Period 1 (July, September, November 2014) and Period 2 (January, March, May 2015)}
\hspace{-2.3cm}    \begin{tabular}{p{3.9cm}rrrrp{3.2cm}rrrr}
    \toprule
    Region 	&	 Index 1	&	 Rank 1	&	Index 2 	&	Rank 2	&	    Region 	&	 Index 1 	&	 Rank 1	&	Index 2	&	Rank 2	\\
    \midrule																	Kabardino-Balkaria	&	311.5	&	1	&	311.5	&	1	&	Krasnoyarsk	&	-18.3	&	42	&	-13.3	&	41	\\
    Nenets AO	&	191.4	&	2	&	268.2	&	3	&	Lipetsk	&	-18.7	&	43	&	-51.3	&	64	\\
    Khanty-Mansi AO	&	156.2	&	3	&	309.8	&	2	&	Nizhny Novgorod	&	-22.8	&	44	&	-31.9	&	51	\\
    Altai (Republic)	&	150.1	&	4	&	100.8	&	6	&	Karachay-Cherkessia	&	-22.8	&	45	&	-46.5	&	61	\\
    Yamalo-Nenets AO	&	145.0	&	5	&	167.9	&	4	&	Penza	&	-24.8	&	46	&	-8.5	&	36	\\
    Buryatia	&	116.2	&	6	&	55.7	&	10	&	Chelyabinsk	&	-25.4	&	47	&	18.6	&	24	\\
    Dagestan	&	86.5	&	7	&	60.0	&	8	&	Chuvashia	&	-25.6	&	48	&	-34.4	&	53	\\
    Astrakhan	&	85.1	&	8	&	-8.8	&	37	&	Saratov	&	-27.8	&	49	&	-4.9	&	35	\\
    Perm	&	83.6	&	9	&	112.4	&	5	&	Sakhalin	&	-28.4	&	50	&	4.3	&	30	\\
    Kaliningrad	&	71.7	&	10	&	19.1	&	23	&	Kurgan	&	-30.1	&	51	&	-26.4	&	48	\\
    Pskov	&	59.9	&	11	&	4.0	&	31	&	Krasnodar	&	-33.3	&	52	&	-15.8	&	42	\\
    Yaroslavl	&	59.5	&	12	&	36.8	&	19	&	Irkutsk	&	-33.3	&	53	&	-73.4	&	74	\\
    Zabaykalsky	&	42.1	&	13	&	37.4	&	18	&	Rostov	&	-33.4	&	54	&	-53.8	&	67	\\
    Ingushetia	&	40.2	&	14	&	49.1	&	12	&	Samara	&	-33.9	&	55	&	-20.9	&	45	\\
    Chukotka AO	&	35.9	&	15	&	50.0	&	11	&	Khakassia	&	-34.4	&	56	&	-51.7	&	65	\\
    Kamchatka	&	34.7	&	16	&	71.2	&	7	&	Karelia	&	-34.5	&	57	&	-43.9	&	60	\\
    Stavropol	&	26.9	&	17	&	-47.3	&	62	&	Kalmykia	&	-35.6	&	58	&	-23.9	&	47	\\
    Jewish	&	19.5	&	18	&	40.0	&	16	&	North Ossetia-Alania	&	-36.0	&	59	&	-34.0	&	52	\\
    Murmansk	&	18.2	&	19	&	37.8	&	17	&	Kursk	&	-36.2	&	60	&	-16.7	&	43	\\
    Bryansk	&	16.1	&	20	&	57.6	&	9	&	Kaluga	&	-39.3	&	61	&	-67.9	&	73	\\
    Leningrad + St.Petersburg	&	11.1	&	21	&	44.1	&	14	&	Tula	&	-41.4	&	62	&	-12.3	&	39	\\
    Tomsk	&	10.4	&	22	&	12.0	&	27	&	Tuva	&	-41.7	&	63	&	-40.5	&	58	\\
    Tyumen	&	9.9	&	23	&	24.2	&	22	&	Oryol	&	-41.9	&	64	&	-40.7	&	59	\\
    Vologda	&	8.6	&	24	&	-39.2	&	57	&	Komi	&	-43.7	&	65	&	-22.9	&	46	\\
    Volgograd	&	8.0	&	25	&	14.2	&	26	&	Magadan	&	-44.1	&	66	&	-59.4	&	71	\\
    Belgorod	&	7.7	&	26	&	44.0	&	15	&	Ryazan	&	-44.3	&	67	&	-34.9	&	56	\\
    Vladimir	&	4.1	&	27	&	-8.9	&	38	&	Mari El	&	-44.5	&	68	&	-76.0	&	75	\\
    Tver	&	1.2	&	28	&	5.7	&	29	&	Tatarstan	&	-44.5	&	69	&	-78.1	&	76	\\
    Omsk	&	-4.7	&	29	&	-31.3	&	50	&	Udmurtia	&	-44.7	&	70	&	-48.9	&	63	\\
    Amur	&	-6.1	&	30	&	-29.0	&	49	&	Adygea	&	-44.9	&	71	&	-66.2	&	72	\\
    Novgorod	&	-7.2	&	31	&	24.9	&	21	&	Chechnya	&	-45.2	&	72	&	-34.7	&	54	\\
    Kostroma	&	-9.4	&	32	&	-13.3	&	40	&	Sakha	&	-48.8	&	73	&	-56.8	&	69	\\
    Ulyanovsk	&	-10.9	&	33	&	14.7	&	25	&	Mordovia	&	-49.6	&	74	&	-58.4	&	70	\\
    Orenburg	&	-11.0	&	34	&	11.3	&	28	&	Arkhangelsk	&	-52.2	&	75	&	-52.2	&	66	\\
    Tambovsk	&	-11.1	&	35	&	25.4	&	20	&	Smolensk	&	-54.1	&	76	&	-56.2	&	68	\\
    Sverdlovsk	&	-11.5	&	36	&	48.1	&	13	&	Voronezh	&	-66.3	&	77	&	-97.3	&	79	\\
    Khabarovsk	&	-11.7	&	37	&	0.6	&	32	&	Ivanovo	&	-72.8	&	78	&	-98.8	&	80	\\
    Novosibirsk	&	-15.1	&	38	&	-18.7	&	44	&	Altai	&	-72.8	&	79	&	-90.7	&	78	\\
    Primorsky	&	-15.3	&	39	&	-34.8	&	55	&	Kemerovo	&	-79.4	&	80	&	-86.2	&	77	\\
    Moscow	&	-17.1	&	40	&	-0.3	&	34	&	Kirov	&	-90.8	&	81	&	-119.5	&	81	\\
    Bashkortostan	&	-18.2	&	41	&	0.0	&	33	&		&		&		&		&		\\
    \bottomrule
    \end{tabular}%
  \label{tab:ranking}}%
\end{table}%

\subsection{Is there an Environmental Kuznets Curve?}
The results of the parameter suppose a curvilinear relationship between GRP per capita and EA of the regions. Nevertheless, we cannot find a structure supporting the Environmental Kuznets Curve hypothesis.

\begin{figure}[!ht]
	\centering
	\includegraphics[trim=0mm 0mm 0mm 0mm, clip, width=0.75\textwidth]{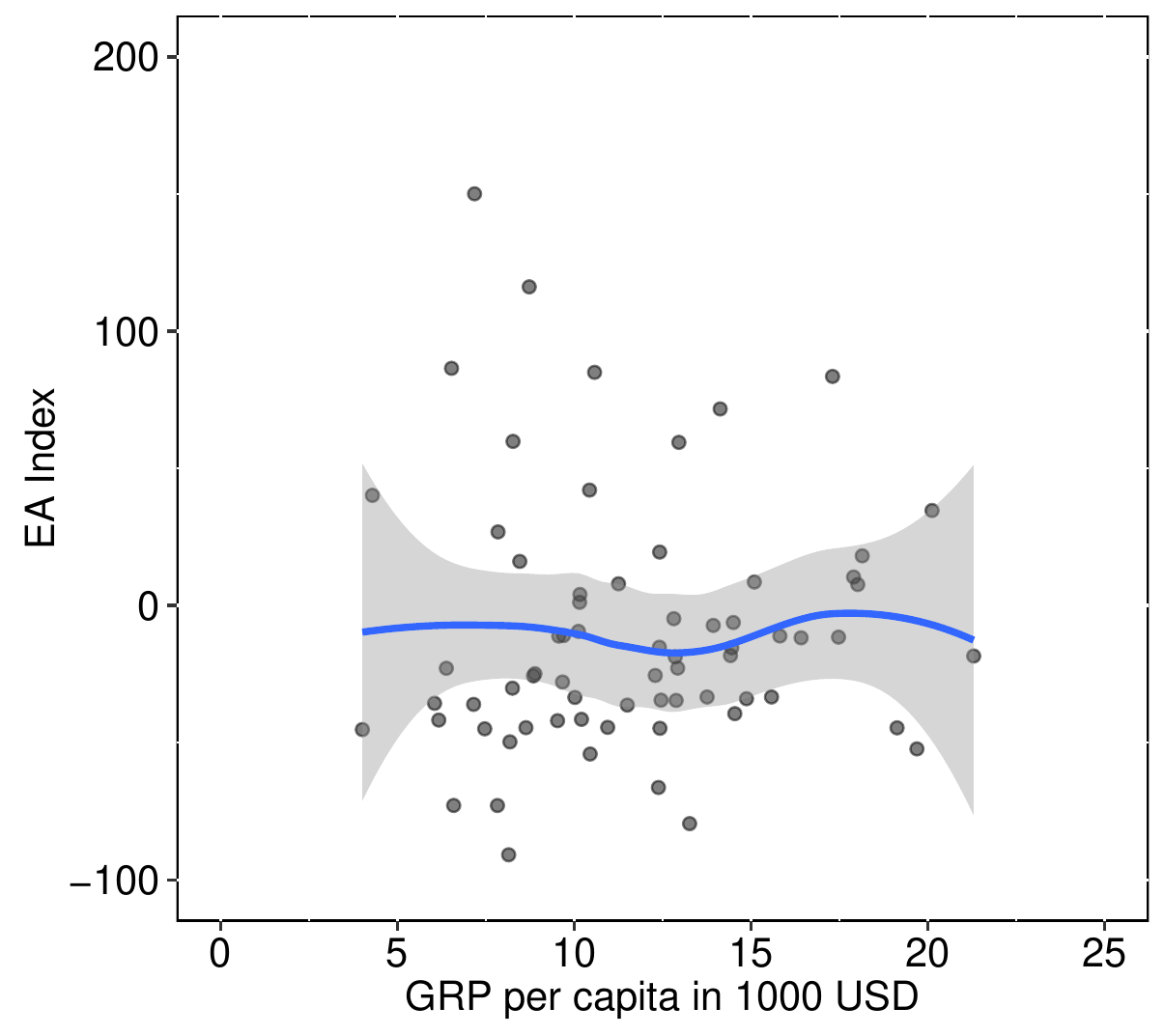}
	\caption{Relationship between GRP per capita in purchasing power parity and the EA Index using local polynomial regression fit (blue), 95\% pointwise confidence interval (grey area).}
	\label{fig:kuz}
\end{figure}

In Figure \ref{fig:kuz} the estimated index values are plotted against the GRP per capita in purchasing power parity of the Russian regions. The curve is fitted by local polynomial regression enclosed by a 95\% pointwise confidence interval. Against our expectations, there is not a relationship, which can be described through the EKC.
One of the reasons could be the subjective selection of more than 200 environmental phrases. Possibly, the number of queries of other environment phrases deliver a different picture.

\newpage
\section{Concluding Remarks}
\label{sec:Concluding}
This paper makes use of the MIMIC model to estimate an index of environmental awareness. The weights assigned to the various indicators are therefore determined endogenously in contrast to the procedure in indices of environmental quality. This seems to be the correct approach for countries with a still emerging EA. 
As there was the sharply RUR devaluation at the end of 2014. We expected a changing in the interest of environment topics of the population in the regions. However, we do not find significant differences between the parameter values of the indicator variables in 2014 and 2015. It is possible, that the RUR devaluation affects the number of queries only after a certain delay. For this reason, we need to analyse query data from the second half year of 2015.

The results also show that the EKC hypothesis cannot hold true. With more data and more appropriate data it should be possible to obtain clearer results. Perhaps the ethnic composition of the regions of the Russian Federation could help to further improve the results of the estimations.

\newpage
\input{appendix}

\end{document}

%% file: appendix.tex
\section*{Appendix}

\begin{figure}[!ht]
\hspace*{-1cm}
	\centering
	\includegraphics[width=1.1\textwidth]{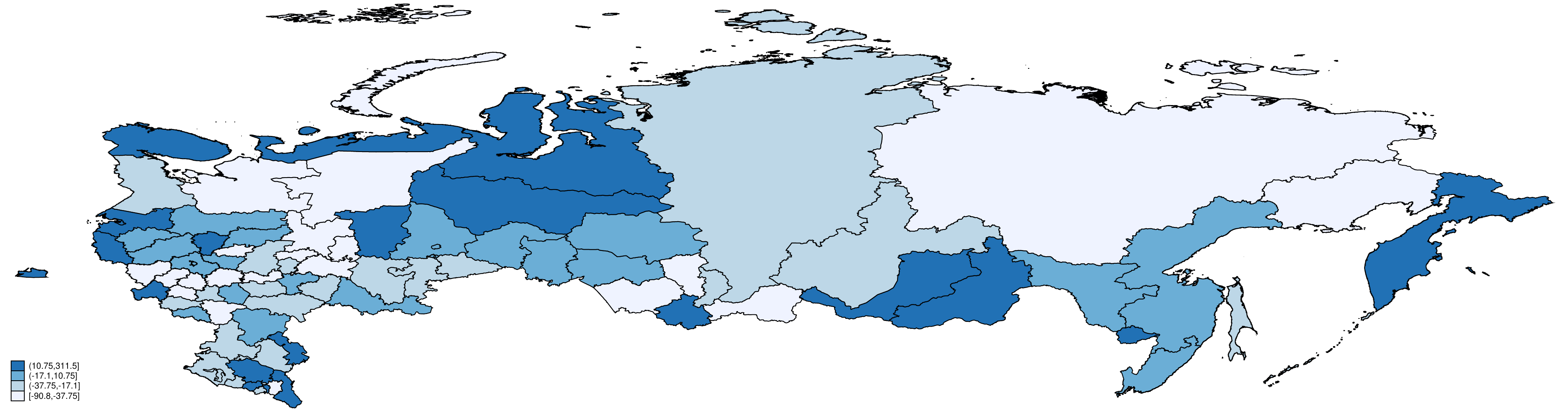}
	\caption{Map of 80 Russian regions illustrating the Environmental Awareness Index for 2014 (without Autonomous Okrug Chukotka). High ranked regions are dark blue and low ranked regions light blue.}
	\label{fig:map2014}
\end{figure}

\vspace{2cm}
\begin{figure}[!ht]
\hspace*{-1cm}
	\centering
	\includegraphics[width=1.1\textwidth]{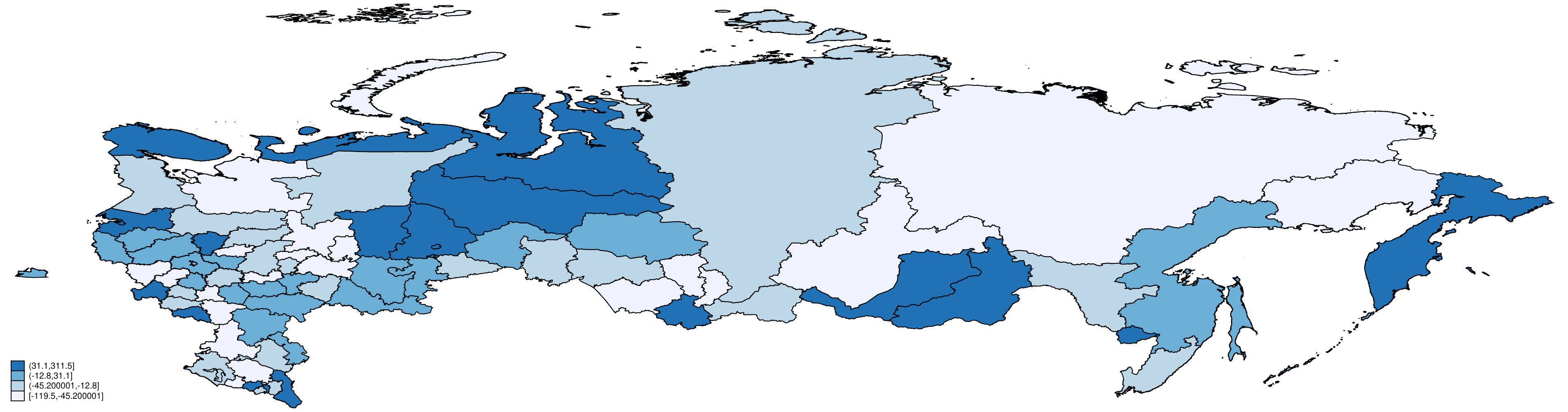}
	\caption{Map of 80 Russian regions illustrating the Environmental Awareness Index for 2015 (without Autonomous Okrug Chukotka). High ranked regions are dark blue and low ranked regions light blue.}
	\label{fig:map2015}
\end{figure}

%% file: EAIndexArxiv.bbl
\newpage

\begin{thebibliography}{10}
\bibitem[Ajzen, 1991]{Ajzen1991} Ajzen, I. (1991) 'The theory of planned behavior', {\itshape Organizational Behavior and Human Decision Processes}, Vol. 50, No. 2, pp. 179-211,
\url{http://dx.doi.org/10.1016/0749-5978(91)90020-T}.
\bibitem[Bollen, 1989]{Bollen1989} Bollen, K. A. (1989) {\itshape Structural Equations with Latent Variables}, John Wiley \& Sons, Inc., New York.
\bibitem[Bollen and Stine, 1992]{Bollen1992} Bollen, K. A. and Stine, R. A. (1992) 'Bootstrapping Goodness-of-Fit Measures in Structural Equation Models', {\itshape Sociological Methods \& Research}, Vol. 21, No. 2, pp. 205--229. 
\bibitem[Bentler and Satorra, 2001]{Bentler2001}
Bentler, P. and Satorra, A. (2001) 'A Scaled Difference Chi-square Test Statistic for Moment Structure Analysis', {\itshape Psychometrika}, Vol. 66, No. 4, pp. 507--514.
\bibitem[Borick et. al. (2011)]{Borick2011} Borick, C. P., Lachapelle, E. and Rabe, B. G. (2011) 'Climate compared: public opinion on climate change in the United States and Canada', {\itshape Brookings Institution Issues in Governance Studies}, Vol. 39.
\bibitem[Brajer et. al., 2011]{Brajer2011} Brajer, V., Mead, R. W. and Xiao, F. (2011) 'Searching for an Environmental Kuznets Curve in China's air pollution', {\itshape China Economic Review}, Vol. 22, pp. 383--397. 
\url{http://dx.doi.org/10.1016/j.chieco.2011.05.001}.
\bibitem[Brown et. al., 1992]{browne1992alternative} Browne, M. W. and Cudeck, R. (1992) 'Alternative ways of assessing model fit', {\itshape Sociological Methods \& Research}, Vol. 21, No. 2, pp. 230--258. 
\bibitem[Buehn and Farzanegan (2013)]{Buehn2013} Buehn, A. and Farzanegan, M. R. (2013) 'Hold your breath: a new index of air pollution', {\itshape Energy Economics}, Vol. 37, pp. 104--113.
\url{http://dx.doi.org/10.1016/j.eneco.2013.01.011}.
\bibitem[Cassou and Hamilton (2004)]{cassou2004} Cassou, S. P. and Hamilton S. F. (2004) 'The transition from dirty to clean industries: optimal fiscal policy and the environmental Kuznets curve', {\itshape Journal of Environmental Economics and Management}, Vol. 48, No. 3, pp. 1050--1077. 
\bibitem[Diederich and Goeschl (2014)]{Diederich2014} Diederich, J., Goeschl, T. (2014) 'Willingness to pay for voluntary climate action and its determinants: field-experimental evidence', {\itshape Environmental and Resource Economics}, Vol. 57, No. 3, pp. 405--429. \url{http://dx.doi.org/10.1007/s10640-013-9686-3}.
\bibitem[Dinda (2004)]{Dinda2004} Dinda, S. (2004) 'Environmental Kuznets Curve Hypothesis: A Survey', {\itshape Ecological Economics}, Vol. 49, pp. 431--455. \url{http://dx.doi.org/10.1016/j.ecolecon.2004.02.011}.
\bibitem[Ekholm et.~al., 2010]{Ekholm2010} Ekholm, T., Soimakallio, S., Höhne, N. Syri, S. and Savolainen, I. (2010) 'Effort sharing in ambitious, global climate change mitigation scenarios', {\itshape Energy Policy}, Vol. 38, No. 4, pp. 1797--1810. \url{http://dx.doi.org/10.1016/j.enpol.2009.11.065}.
\bibitem[European Commission (2008)]{EC2008} European Commission (2008) {\itshape 'Europeans' attitudes towards climate change. Special Eurobarometer 300} [online], European Parliament / European Commission. \url{http://ec.europa.eu/public_opinion/archives/ebs/ebs_300_full_en.pdf}. (Accessed October 2015)
\bibitem[Fosten et al. (2012)]{Fosten2012} Fosten, J., Morley, B. and Taylor, T. (2012) 'Dynamic misspecification in the environmental Kuznets curve: Evidence from CO$_2$ and SO$_2$ emissions in the United Kingdom', {\itshape Ecological Economics}, Vol. 76, pp. 25--33. \url{http://dx.doi.org/10.1016/j.ecolecon.2012.01.023}.
\bibitem[Goldberger and Hauser (1971)]{Goldberger1971} Goldberger, A. S. and Hauser, R. (1971) 'The treatment of unobservable variables in path analysis', {\itshape Sociological methodology}, Vol. 3, No. 8, pp. 1--8. 
\bibitem[Golub et. al. (2013)]{Golub2013} Golub, A., Kozeltsev, M., Martusevich, A. and Strukova, E. (2013) 'The challenge of reforming environmental regulations in Russia', In Alexeev, M. and Weber, S. (Eds.), {\itshape The Oxford Handbook of the Russian Economy}, Oxford University Press, ISBN 978-0-19-975992-7, pp. 426--450.
\bibitem[Grossman and Krueger (1995)]{Grossman1995} Grossman, G. M. and Krueger, A. B. (1995) 'Economic growth and the environment', {\itshape Quarterly Journal of Economics}, Vol. 110, pp. 353--377.
\bibitem[He and Richard, 2010]{He2010} He, J. and Richard, P. (2010) 'Environmental Kuznets curve for CO2 in Canada', {\itshape Ecological Economics}, Vol. 69, pp. 1083--1093. \url{http://dx.doi.org/10.1016/j.ecolecon.2009.11.030}.
\bibitem[Huang et. al. (2008)]{Huang2008} Huang, W. M., Lee, G. W. M. and Wu, C. C. (2008) 'GHG emissions, GDP growth and the Kyoto Protocol: A revisit of Environmental Kuznets Curve hypothesis', Figure~3, {\itshape Energy Policy}, Vol. 36, pp. 239--247. \url{http://dx.doi.org/10.1016/j.enpol.2007.08.035}.
\bibitem[International Energy Agency (IEA) Statistics (2013)]{IEA2013} International Energy Agency (IEA) Statistics (2013) {\itshape CO$_2$ Emissions from Fuel Combustion} [online], Paris, France. \url{http://www.iea.org/publications/freepublications/publication/name,43840,en.html}. (Accessed August 2014)
\bibitem[Intergovernmental Panel on Climate Change (IPCC), 2014]{IPCC2014} Intergovernmental Panel on Climate Change (IPCC) (2014) {\itshape Climate Change 2014: Impacts, Adaptation, and Vulnerability} [online],  Cambridge University Press, Cambridge and New York, chapter 24. 
\url{http://ipcc-wg2.gov/AR5/images/uploads/WGIIAR5-Chap24_FGDall.pdf}. (Accessed August 2014)
\bibitem[J\"oreskog and Goldberger (1975)]{Joereskog1975} J{\"o}reskog, K. G. and Goldberger, A. S. (1975) 'Estimation of a model with multiple indicators and multiple causes of a single latent variable', {\itshape Journal of the American Statistical Association}, Vol. 70, No. 315a, pp. 631--639 .
\bibitem[Karytsas and Theodoropoulou (2004)]{Karytsas2014} Karytsas, S., Theodoropoulou, H. (2014) 'Socioeconomic and demographic factors that influence publics' awareness on the different forms of renewable energy sources', {\itshape Renewable Energy}, Vol. 71, pp. 480--485. 
\url{http://dx.doi.org/10.1016/j.renene.2014.05.059}.
\bibitem[Kline, 2011]{Kline2011} Kline, R. B. (2011), {\itshape Principles and practice of structural equation modeling}, 3rd ed., The Guilford Press, London.
\bibitem[List and Gallet, 1999]{List1999} List, J. A. and Gallet, C. A. (1999) 'The environmental Kuznets curve: Does one size fit all?', {\itshape Ecological Economics}, Vol. 31, No. 3, pp. 409--423.
\bibitem[Lorenzoni and Pidgeon (2006)]{Lorenzoni2006} Lorenzoni, I. and Pidgeon, N. (2006) 'Public views on climate change: European and USA perspectives', {\itshape Climatic Change}, Vol. 77, No. 1, pp. 73--95.
\bibitem[Maddison (2006)]{maddison2006} Maddison, D. (2006) 'Environmental Kuznets curves: A spatial econometric approach', {\itshape Journal of Environmental Economics and Management}, Vol. 51, No. 2, pp. 218--230. 
\bibitem[MacCallum et.~al., 1996]{maccallum1996} MacCallum, R. C. and Browne, M. W. and Sugawara, H. M. (1996) 'Power analysis and determination of sample size for covariance structure modeling', {\itshape Psychological methods}, Vol. 1, No. 2, pp. 130--149. 
\bibitem[Olsen et. al., 2000]{Olson2000} Olsson, U. H., Foss, T., Troye, S. V. and Howell, R. D. (2000) 'The Performance of ML, GLS, and WLS Estimation in Structural Equation Modeling Under Conditions of Misspecification and Nonnormality', {\itshape Structural Equation Modeling: A Multidisciplinary Journal}, Vol. 7, No. 4, 557--595.
\bibitem[Rabe-Hesketh et. al., 2008]{Rabe-Hesketh2008} Rabe-Hesketh, S., Skrondal, A. and Zheng, X. (2008) 'Multilevel structural equation modeling', in Lee, S. (Eds.), {\itshape Handbook on Structural Structural Equation Models}, Elsevier, Amsterdam,  pp. 209--227.
\bibitem[Robinson, 1974]{Robinson1974} Robinson, P. (1974) 'Identification, estimation, and large sample theory for regressions containing unobservable variables', {\itshape International Economic Review}, Vol. 15, No. 3, pp. 680--692.
\bibitem[Satorra and Bentler (1994)]{Satorra1994} Satorra, A. and Bentler, P.M. (1994) 'Corrections to test statistics and standard errors in covariance structure analysis', in von Eye, A.  and Clogg, C. C. (Eds.), {\itshape Latent Variables Analysis: Applications for Developmental Research Sage},
Thousand Oaks, CA, pp. 399--419.
\bibitem[Selden and Song, 1994]{Selden1994} Selden, T. M. and Song, D. (1994) 'Environmental quality and development: Is there a Kuznets curve for air pollution emissions?', {\itshape Journal of Environmental Economics and Management}, Vol. 27, No. 2, pp. 147--162.
\bibitem[Soyez et. al. (2009)]{Soyez2009} Soyez, K., Hoffmann, S., Wünschmann, S. and Gelbrich, K. (2009) 'Proenvironmental value orientation across cultures', {\itshape Social Psychology}, Vol. 40, No. 4, pp. 222--233. \url{http://dx.doi.org/10.1027/1864-9335.40.4.222}.
\bibitem[Soyez (2012)]{Soyez2012} Soyez, K. (2012) 'How national cultural values affect pro-environmental consumer behavior', {\itshape International Marketing Review}, Vol. 29, No. 6, pp. 623--646. 
\url{http://dx.doi.org/10.1108/02651331211277973}.
\bibitem[Stern (2004)]{Stern2004} Stern D. I. (2004) 'The Rise and Fall of the Environmental Kuznets Curve', {\itshape World Development}, Vol. 32, No. 8, pp. 1419--1439.
\bibitem[Trujillo-Ortiz and Hernandez-Wall ( 2003)]{Trujillo-Ortiz2003} Trujillo-Ortiz, A. and Hernandez-Walls, R. (2003), 'Mskekur: Mardia's multivariate skewness and kurtosis coefficients and its hypotheses testing'. A MATLAB File.
\url{http://www.mathworks.com/matlabcentral/fileexchange/loadFile.do?objectId=3519}. (Access August 2014)
\bibitem[United Nations Development Programme (2013)]{UNDP2013} UNDP (2013) {\itshape National Human Development Report for the Russian Federation 2013} [online],  Sustainable Development:
Rio Challenges, LLC RS Ilf, Rio, pp. 48--139.
\url{http://www.undp.ru/index.php?lid=1&cmd=publications1&id=48}. (Access September 2014)
\bibitem[Wang, 2013]{Wang2013} Wang, Y.-C. (2013) 'Functional sensitivity of testing the environmental Kuznets curve hypothesis', {\itshape Resource and Energy Economics}, Vol. 35, pp. 451--466. 
\url{http://dx.doi.org/10.1016/j.reseneeco.2013.01.003}.
\bibitem[Weber and Wiesmeth (2015)]{Weber2015} Weber, S. and Wiesmeth, H. (2015) 'The Environmental Kuznetsk Curve: Awareness of climate change', {\itshape Discussion Paper}, TU Dresden.
%\bibitem{Wiesmeth2011} Wiesmeth H (2011): Environmental Economics: Theory and Policy in Equilibrium. Springer, Berlin.
\bibitem[Yang et. al., 2015]{Yang2015} Yang, H. and He, J. and Chen, S. (2015) 'The fragility of the Environmental Kuznets Curve: Revisiting the hypothesis with Chinese data via an ``Extreme Bound Analysis'' ', {\itshape Ecological Economics}, Vol. 109, pp. 41--58  \url{http://dx.doi.org/10.1016/j.ecolecon.2014.10.023}.
\bibitem[Yuan and Bentler (2000)]{Yuan2000} Yuan, K. and Bentler, P. M. (2000) 'Three likelihood-based methods for mean and covariance structure analysis with nonnormal missing data', {\itshape Sociological methodology}, Vol. 30, pp. 165--200.
\bibitem[Zyadin et. al. (2014)]{Zyadin2014} Zyadin A., Halder, P., Kähkönen T. and Puhakka, A. (2014) 'Challenges to renewable energy: A bulletin of perceptions from international academic arena', {\itshape Renewable Energy}, Vol. 69, pp. 82-88. \url{http://dx.doi.org/10.1016/j.renene.2014.03.029}.
\bibitem[OECD (2016)]{OECD} Organisation for Economic Co-operation and Development (2016), {\itshape OECD.Stat public database} [online], \url{http://stats.oecd.org/Index.aspx?datasetcode=REG_DEMO_TL2#}. (Access January 2016)
\bibitem[RFSSS (2016)]{RFSSS} Federal Statistics Service of Russia (2016), {\itshape RFSSS public database} [online], \url{http://www.gks.ru/wps/wcm/connect/rosstat_main/rosstat/ru/}. (Access January 2016)
\bibitem[The World Bank (2016)]{WorldB} The World Bank (2016), {\itshape Worldbank public database} [online], \url{http://search.worldbank.org/}. (Access January 2016)
\end{thebibliography}
